\documentclass[twoside,12pt]{article}
\usepackage{indentfirst}
\usepackage{bm}
\usepackage{epsfig}
\usepackage{graphicx}
\usepackage{cite}
\usepackage{dcolumn}
\usepackage[colorlinks=true,linkcolor=blue,anchorcolor=blue,citecolor=blue,urlcolor=blue]{hyperref}
\usepackage{amssymb}
\usepackage{amsmath}
\usepackage{cleveref}
\usepackage{soul}

\begin{document}

\thispagestyle{empty} \vspace*{0.8cm}\hbox
to\textwidth{\vbox{\hfill\huge\sf Commun. Theor. Phys.\hfill}}
\par\noindent\rule[3mm]{\textwidth}{0.2pt}\hspace*{-\textwidth}\noindent
\rule[2.5mm]{\textwidth}{0.2pt}


\begin{center}
\LARGE\bf The synthetic gauge field and exotic vortex phase with spin-orbital-angular-momentum coupling
\end{center}

\footnotetext{\hspace*{-.45cm}\footnotesize $^*$E-mail:dengyg3@mail.sysu.edu.cn}
\footnotetext{\hspace*{-.45cm}\footnotesize $^\&$These authors contributed equally to this work }

\begin{center}
\rm Yingqi Liu$^{\rm 1)\&}$ , \ \ Yun Chen$^{\rm 1)\&}$, \ and  \ Yuangang Deng$^{\rm 1)*}$
\end{center}

\begin{center}
\begin{footnotesize} \sl
	Guangdong Provincial Key Laboratory of Quantum Metrology and Sensing \&\\
	School of Physics and Astronomy, Sun Yat-Sen University(Zhuhai Campus), Zhuhai 519082, China, \\

${}^{\rm 1)}$ \\ 
\end{footnotesize}
\end{center}

\begin{center}

\end{center}

\vspace*{2mm}

\begin{center}
\begin{minipage}{15.5cm}
\parindent 20pt
\footnotesize
Ultracold atoms endowed with tunable spin-orbital-angular-momentum coupling (SOAMC) represent a promising avenue for delving into exotic quantum phenomena. Building on recent experimental advancements, we propose the generation of synthetic gauge fields ,and by including exotic vortex phases within spinor Bose-Einstein condensates, employing a combination of a running wave and Laguerre-Gaussian laser fields. We investigate the ground-state characteristics of the SOAMC condensate, revealing the emergence of exotic vortex states with controllable orbital angular momenta. It is shown that the interplay of the SOAMC and conventional spin-linear-momentum coupling induced by the running wave beam leads to the formation of a vortex state exhibiting a phase stripe hosting single multiply quantized singularity. The phase of the ground state will undergo the phase transition corresponding to the breaking of rotational symmetry while preserving the mirror symmetry. Importantly, the observed density distribution of the ground-state wavefunction, exhibiting broken rotational symmetry, can be well characterized by the synthetic magnetic field generated through light interaction with the dressed spin state. Our findings pave the way for further exploration into the rotational properties of stable exotic vortices with higher orbital angular momenta against splitting in the presence of synthetic gauge fields in ultracold quantum gases. 
\end{minipage}
\end{center}

\begin{center}
\begin{minipage}{15.5cm}
\begin{minipage}[t]{2.3cm}{\bf Keywords:}\end{minipage}
\begin{minipage}[t]{13.1cm}
ultracold atoms, synthetic gauge fields, vortex phases, spin-orbital-angular-momentum coupling
\end{minipage}\par\vglue8pt

\end{minipage}
\end{center}

\section{Introduction}\label{one}

Ultracold quantum gases, renowned for their unparalleled controllability, are opening up new frontiers for the exploration of novel quantum simulators~\cite{RevModPhys.80.885,RevModPhys.85.1191,Cirac+2021+453+456,Bloch2012}. The advent of synthetic gauge fields for ultracold neutral atoms has unveiled unprecedented opportunities for the exploration of novel topological quantum matters~\cite{RevModPhys.83.1523,Goldman_2014,Zhai_2015,PhysRevLett.95.010404,PhysRevLett.102.130401,RevModPhys.82.3045,RevModPhys.83.1057}. Unlike the simulation of effective Lorentz forces in rotating neutral atoms~\cite{doi:10.1080/00018730802564122,RevModPhys.81.647,PhysRevLett.84.806,PhysRevLett.92.050403,PhysRevLett.92.040404,doi:10.1126/science.1060182,doi:10.1126/science.aba7202}, the application of light-induced synthetic gauge fields that target the dressed spin state enables the creation of a giant artificial magnetic field, thereby broadening the scope of applications in quantum science and technology~\cite{RevModPhys.83.1523,Goldman_2014,Zhai_2015,PhysRevLett.95.010404,PhysRevLett.102.130401}. Among various implementations, the employment of optical Raman coupling processes and degenerate dark states has been experimentally realized in both bosonic and fermionic atomic species~\cite{Lin2011,li2017stripe,PhysRevLett.109.095302,huang2016experimental,doi:10.1126/science.aaf6689}. By utilizing spatially dependent Raman coupling with internal spin states of atoms, various types of spin-orbit coupling (SOC) can emerge~\cite{doi:10.1126/science.aao5392}, facilitating the transfer of photon momentum to the atom with the change of the internal spin states. This SOC interaction between atomic spin and the center-of-mass motion of atoms, known as spin-linear-momentum coupling (SLMC)~\cite{Lin2011,li2017stripe,PhysRevLett.109.095302,huang2016experimental,doi:10.1126/science.aaf6689,PhysRevLett.102.046402,PhysRevLett.108.225301,PhysRevA.97.053609}, has garnered significant interest.

In recent years, the coupling of atomic spin to a Laguerre-Gaussian (LG) Raman beam, which carries orbital angular momentum (OAM), has attracted great interest ~\cite{PhysRevX.2.041011,RevModPhys.89.035004,Clark_2012,PhysRevLett.108.044801}. This new form of SOC, named spin-orbital-angular-momentum coupling (SOAMC), has provided a new paradigm for the exploration of a variety of intriguing vortex
states and rich quantum phenomena in both spinor Bose Einstein condensates (BECs) and Fermi gases~\cite{PhysRevLett.121.113204,PhysRevA.91.063627,PhysRevLett.122.110402,PhysRevA.91.033630,PhysRevA.93.013629,PhysRevA.94.033627,PhysRevA.102.063328,PhysRevA.105.063308,PhysRevLett.125.260407,PhysRevLett.126.193401,PhysRevA.108.043310,PhysRevA.105.023320}.  Of particular interest, SOAMC possesses spatial rotational symmetry and the breaking of translational symmetry significantly influences strongly correlated physics in atomic gases, e.g. with respect to the diverse spectrum properties~\cite{PhysRevLett.122.110402} and topological excitations~\cite{PhysRevResearch.2.033152,PhysRevA.92.033615}. Moreover, ultracold atoms mediated by SOAMC, possessing an azimuthal gauge potential, can be utilized to measure superfluid fractions~\cite{PhysRevLett.104.030401}, study artificial magnetic fields and investigate monopoles~\cite{PhysRevLett.121.250401,PhysRevLett.108.035301,PhysRevLett.90.140403,Ray2014,doi:10.1126/science.1258289}. The interplay between SOAMC and SLMC invites exploration into exotic vortex phases and synthetic gauge fields, potentially enriching our understanding SOC-mediated novel quantum matters and inspiring further experimental investigations in the ultracold atoms community.

In this work, we demonstrate the realization of SLMC and SOAMC in a spin-$1/2$ condensate using a Raman process facilitated by a combination of a running wave and LG laser fields. The intensity of SLMC can be modulated by adjusting the relative angle between the Raman beams. Our investigation into the ground-state phases of the condensate, mediated by these two types of SOC, reveals the formation of exotic vortex states with controllable orbital angular momenta. We find that the interplay of the SLMC and SOAMC gives rise to the novel vortex state with a phase stripe that encompasses a single, multiply quantized singularity with the rotational symmetry breaking. Furthermore, the density distribution associated with the broken rotational symmetry can be effectively interpreted through the synthetic magnetic field created by light acting on the dressed spin state; notably, the synergy mirror symmetry in both the condensate density distribution and the synthetic gauge field.

This paper is organized as follows. In section~\ref{two}$\quad$, we discuss the design of the model and the Hamiltonian. In section~\ref{three}$\quad$, we present the synthetic magnetic field and condensate density distribution of the ground state resulting from the SOAMC. In section~\ref{four}$\quad$, we study the exotic vortex phase emerging as a result of the combined effects of  SOAMC and SLMC. Finally, we provide concluding remarks in section~\ref{five}$\quad$

\section{Model and Hamiltonian}\label{two}
\color{red}
\begin{figure}[h]
	\centering
	\includegraphics[width=0.65\textwidth]{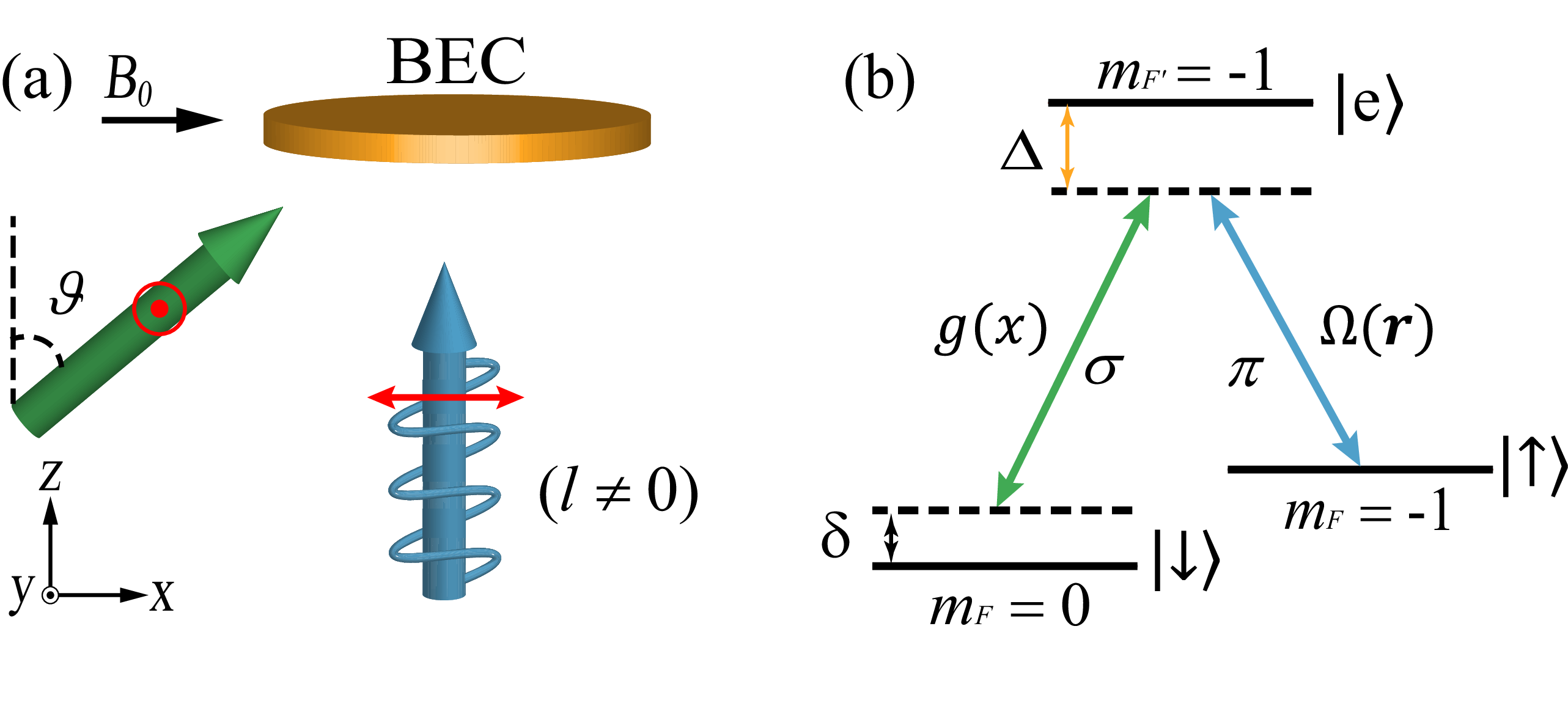}
	\caption{(a) A scheme of a spinor condensate for creating both SLMC and SOAMC by using a $\sigma$-polarized running wave laser and a $\pi$-polarized LG laser carrying nonzero OAM ($l\neq0$).Here, ${\vartheta}$ is an introduced knob by controlling the strength of SLMC. (b) An energy level diagram.}
	\label{Fig1}
\end{figure}
\color{black}
We consider a BEC comprising $N$ atoms, confined within a quasi two-dimensional circular optical box trap~\cite{PhysRevLett.110.200406,Navon2016}. Figure~\ref{Fig1}(a) illustrates the laser configuration for creating SOC and figure~\ref{Fig1}(b) displays the atomic level structure for the generation of two-photon Raman process. Within the $F=1$ ground electronic manifold of $^{87}$Rb~\cite{Lin2011,PhysRevLett.121.113204,PhysRevLett.122.110402}, two hyperfine states are labeled as $|\uparrow  \rangle=|F=1, m_\mathrm{F}=-1\rangle$ and $|\downarrow  \rangle=|F = 1, m_\mathrm{F} =0\rangle$, defined relative to the Zeeman shift $\hbar \omega_\mathrm{Z}$ produced by a large bias magnetic field $B_0$ along the $x$ axis, which also defines the quantization axis.
The Zeeman shift $\hbar \omega_\mathrm{Z}$ is between $|\uparrow\rangle$ and $|\downarrow\rangle$, whose magnetic quantum numbers of the electronic states satisfy $m_{\uparrow}=m_{\downarrow}-1$. To realize the Raman coupling, an excited state $|\mathrm{e}\rangle=|F'=1, m_\mathrm{F}'=-1\rangle$ is selected to satisfy the selection rules for electric dipole transitions. Specifically, the  linear $\sigma$-polarized (along the $y$ axis) running wave laser can be decomposed as $\sigma^+$ and $\sigma^-$, where the $\sigma^-$ polarized light drives the atomic transition between $|\downarrow\rangle \leftrightarrow | \rm{e} \rangle$ with the magnetic quantum number satisfying $|\Delta m| = 1$. Meanwhile, the linear $\pi$-polarized (along the $x$ axis) LG beam couples the spin-up state $|\uparrow\rangle$ to the excited state $|\rm{e} \rangle$ with the magnetic quantum number satisfying $\Delta m = 0$, as illustrated in figure~\ref{Fig1}(b). As can be seen, the running wave laser is propagating in the direction $\sin\vartheta \bm{\hat{e_x}}+\cos\vartheta \bm{\hat{e_z}}$, where $\vartheta$ is the tunable angle with respect to the $z$ axis. 
The Rabi frequency associated with the running wave laser is denoted as $g(x)=g_0 \mathrm{e}^{\mathrm{i} k_{\mathrm L}\sin\vartheta x}$ with $k_{\mathrm L}=2\pi/\lambda$ being the wave vector of the laser field and $\lambda$ being the wavelength. To establish SOAMC, the atomic transition between $|\uparrow\rangle\leftrightarrow |\rm{e}\rangle$ is linked with a LG laser, characterized by a Rabi frequency $\Omega(r)=\Omega_{0}(r/w_{0})^{l} \mathrm{e}^{\mathrm{i} l\varphi}$ propagating along the $z$ axis. Here, $w_0$ indicates the beam's waist width and the phase winding $\mathrm{e}^{\mathrm{i}l\varphi}$ mirrors the OAM-$\hbar l$ carried by the beam. 
 
Under the conditions of a large atom-pump detuning limit $\Delta\gg\{g_0 ,\Omega_0\}$, the excited atomic state $|\mathrm{e}\rangle$ can be adiabatically eliminated, leading to the derivation of the many-body Hamiltonian for the condensate,
\begin{equation}
	\begin{split}
		\hat{H}=&{\sum_{\sigma\sigma^{\prime}}\int{\mathbf{d}\bm{r}{\hat{\Psi}}_{\sigma}^{\dag}(r)[\hat{h}_{\sigma\sigma^{\prime}}+V(r)\delta_{\sigma\sigma^{\prime}}]{\hat{\Psi}}_{\sigma^{\prime}}(r)}}\\
		&+{\sum_{\sigma\sigma^{\prime}}g_{\sigma\sigma^{\prime}}\int{\mathbf{d}\bm{r}{\hat{\Psi}}_{\sigma}^{\dag}(r){\hat{\Psi}}_{\sigma^{\prime}}^{\dag}(r){\hat{\Psi}}_{\sigma^{\prime}}(r){\hat{\Psi}}_{\sigma}(r)}},
	\end{split}
\end{equation}
where ${\hat{\Psi}}_{\sigma={\uparrow},{\downarrow}}$ denotes the annihilation bosonic atomic field operator for a spin-$\sigma$ atom, $V(r)=V_{\rm{T}}+V_{\rm{LG}}$ is the effective trapping potential with $V_{\rm{T}}$ being the external box trap and $V_{\rm{LG}} = -{\Omega_0^2 r^{2l}}/{(w_0^{2l} \Delta)}$ being the optical Stark shift of the LG laser beam~\cite{PhysRevLett.92.050403}, and $g_{\sigma\sigma'}=4 \pi \hbar^2 a_{\sigma\sigma'}/M$ is the two-body contact interaction with $a_{\sigma\sigma'}$ being the s-wave scattering lengths and $M$ being the mass of an atom. We should note that the constant optical Stark shift of the running wave beam can be neglected.
Additionally, the effective single-particle Hamiltonian, which encompasses both the kinetic energy and the light-induced interactions, is provided by 
\begin{equation}\label{eq2}
\hat{h}=\frac{\bm{p}^2}{2M}\hat{I}+\hbar{\left(\begin{array}{cc}
			\delta/{2} & \Omega r^{l}\mathrm{e}^{\mathrm{i}(k_{\mathrm L}\sin\vartheta x-l\varphi)}\\
			\Omega r^{l}\mathrm{e}^{-\mathrm{i}(k_{\mathrm L}\sin\vartheta x-l\varphi)}& -\delta/{2}
		\end{array}\right)},
\end{equation}
where $\Omega =\Omega_\mathrm{R}/w_0^l$ represents the effective two-photon Raman coupling, with $\Omega_\mathrm{R}=-{g_0 \Omega_0}/{\Delta}$ being the Rabi frequency that relates to light intensities of the Raman lasers. As a result, both SLMC and SOAMC can be generated by tuning the LG laser-mediated Raman process and the relative angle $\vartheta$, corresponding to an interesting light-induced vector potential $\bm{A}$~\cite{RevModPhys.83.1523}. As we will demonstrate, the interplay of these two types of SOC is pivotal for engendering exotic vortex phases and synthetic magnetic fields, with or without rotational symmetry.
In the thermodynamic limit, we adopt the mean-field approach to investigate the ground states of LG laser coupled condensate system in the weakly interacting regime. Within the mean-field framework, we replace the atomic field operators  $\hat{\Psi}_\sigma$ by their mean-field values $\Psi_\sigma=\langle \hat{\Psi}_\sigma\rangle$ in our numerical simulation. To proceed further, the Gross-Pitaevskii equations for the atomic condensate wavefunctions satisfy
\begin{equation}\label{eq3}
\begin{split}
		\mathrm{i} \dot{{\Psi}}_{\uparrow} =& [\frac{{\bm p}^2}{2M}+ V(r)+\frac{\delta}{2}]{\Psi}_{\uparrow}\\
		&+\Omega r^{l}\mathrm{e}^{\mathrm{i}(k_{\mathrm L}\sin\vartheta x-l\varphi)}{\Psi}_{\downarrow}+\mathcal{L}_{\uparrow}{\Psi}_{\uparrow}, \\
		\mathrm{i} \dot{{\Psi}}_{\downarrow} =& [\frac{{\bm p}^2}{2M}+ V(r)-\frac{\delta}{2}]{\Psi}_{\downarrow}\\
		&+\Omega r^{l}\mathrm{e}^{-\mathrm{i}(k_{\mathrm L}\sin\vartheta x-l\varphi)}{\Psi}_{\uparrow}+\mathcal{L}_{\downarrow}{\Psi}_{\downarrow},		
\end{split}
\end{equation}
where $\mathcal{L}_{\sigma} \equiv g_{\sigma\sigma}n_{\sigma}+g_{\uparrow\downarrow}n_{{\sigma'}}$ with density $n_{\sigma}=|{\Psi}_{\sigma}|^2$ for the spin index ${\sigma}\neq{{\sigma}'}$.

Now, we study ground-state configurations of the condensate by solving the Gross-Pitaevskii equations in equation~\eqref{eq3} via imaginary time evolution by numerically minimizing the free energy function $\langle\hat{H}\rangle$. Specifically, we assume a condensate of $N_a =10^5$  $^{87}$Rb BEC, initially prepared in the $| \uparrow \rangle$ state confined in a quasi two-dimensional circular optical box trap~\cite{PhysRevLett.110.200406,Navon2016}. 
The setup features a single-photon recoil energy of $E_{\mathrm L}/\hbar =3.53$kHz(2$\pi$) based on an atomic transition wavelength $\lambda = 803.2$ nm. The $s$-wave scattering lengths for Rb atoms are  $a_{\downarrow \downarrow} = a_{\uparrow \downarrow} =a_{\uparrow \uparrow}= 100.4 a_{\mathrm B}$, with $a_{\mathrm B}$ being the Bohr radius. For simplicity, the collisional interactions are assumed spin-independent under SU(2) symmetry. This yields the typical interaction energy in the tens of Hertz range. We remark that the spin structures for vortex phases presented below is robust against slight variations in the $s$-wave scattering length as short-range contact interaction conserves the number of atoms in the individual spin state.

\section{SOAMC-mediated synthetic magnetic field}\label{three}
For the relative angle  $\vartheta =0$, an interesting SOAMC can be created by the LG laser-mediated two-photon Raman process. In this configuration, the system's the single-particle Hamiltonian $\hat{h}$ maintains rotational symmetry. The SOAMC could give rise to a synthetic magnetic field with different spatial patterns. To better understand the effect of SOAMC and the synthetic magnetic field it generates, we look into the diagonalization of equation~(\ref{eq2}) at a given position ${\bm r}$. This procedure results in two non-degenerate eigenvalues $\lambda_{\pm}/\hbar=\pm {\sqrt{\delta^2+4(\Omega r^l)^2}}/{2}$,  corresponding to the eigenstates $|\chi_{\pm}(x,y) \rangle$ satisfying
\begin{equation}\label{eq4}
	\begin{split}
		|\chi_{-}(x,y)\rangle&=\left(
		\begin{matrix}
			-\cos{\beta} \mathrm{e}^{-\mathrm{i} l \varphi}\\
			\sin{\beta}
		\end{matrix}
		\right),\\
		|\chi_{+}(x,y)\rangle&=\left(
		\begin{matrix}
			\sin{\beta} \mathrm{e}^{-\mathrm{i} l \varphi}\\
			\cos{\beta}
		\end{matrix}
		\right).
	\end{split}
\end{equation}
Here, the mixing angle satisfies the condition
$\tan{\beta}={2\Omega  r^l}/({\sqrt{\delta^2+4(\Omega r^l)^2}-\delta})$. 

Under the adiabatic approximation, the full quantum state of the atom describing both internal and motional degrees of freedom can be expanded in terms of the lower-energy dressed states $|\chi_{-}(x,y)\rangle$. We should note that the adiabatic approximation is valid since the energy gap between the two dressed states ($={\sqrt{\delta^2+4(\Omega r^l)^2}} \geq |\delta|$) is large enough in our simulation\cite{RevModPhys.83.1523}. For its motional degree of freedom, the vector gauge potential $\bm{A}={A}_x \bm{\hat{e}_y}+{A}_y \bm{\hat{e}_y}$ is determined by the spatial derivatives of the dressed state
\begin{equation}\label{eq5}
	\begin{split}
		{A}_{x}&=\mathrm{i} \hbar \langle \chi_{-}(x,y)|\partial_x\chi_{-}(x,y) \rangle\\
		&=\hbar l\cos^2{\beta} \,  \frac{\partial\varphi}{\partial x}, \\
		{A}_{y}&=\mathrm{i} \hbar  \langle  \chi_{-}(x,y)|\partial_y\chi_{-}(x,y)  \rangle \\
		&=\hbar l\cos^2{\beta} \, \frac{ \partial\varphi}{\partial y},
	\end{split}
\end{equation}
Clearly, the vector gauge potential ${\bm A}$ is dominated by the angular quantum number $l$ for the LG beam. According to the spatially dependent vector potential $\bm{A}$ in equation~\eqref{eq5}, the light-induced synthetic magnetic field ${\cal B} = \nabla \times \bm{A} $ takes the form as 
\begin{equation} 
	\begin{split}
		{\cal B}  (\vartheta=0)=&\frac{\hbar}{2}[l(\frac{\partial(\cos{2\beta})}{\partial x}                 \frac{\partial\varphi}{\partial y}\\
		&-\frac{\partial(\cos{2\beta})}{\partial y}  \frac{\partial\varphi}{\partial x})] \bm{\hat{e}_z} \\
		=& \frac{ \delta}{|\delta|}\frac{ \hbar  C l^2 r^{2l-2}}{2 (1+ C r^{2l})^{\frac{3}{2}}} \bm{\hat{e}_z}
	\end{split}
	\label{eq6}
\end{equation} 
with $C = {4 \Omega^2}/{\delta^2}$.
We find that the direction of the synthetic magnetic field is  along the positive (negative) $z$ axis for the two-photon detuning $\delta > 0$ ($<0$). The magnitude of ${\cal B}(\vartheta=0)$ is obviously dependent on the value of $l$.
\begin{figure}[h]
	\centering
	\includegraphics[width=0.58\textwidth]{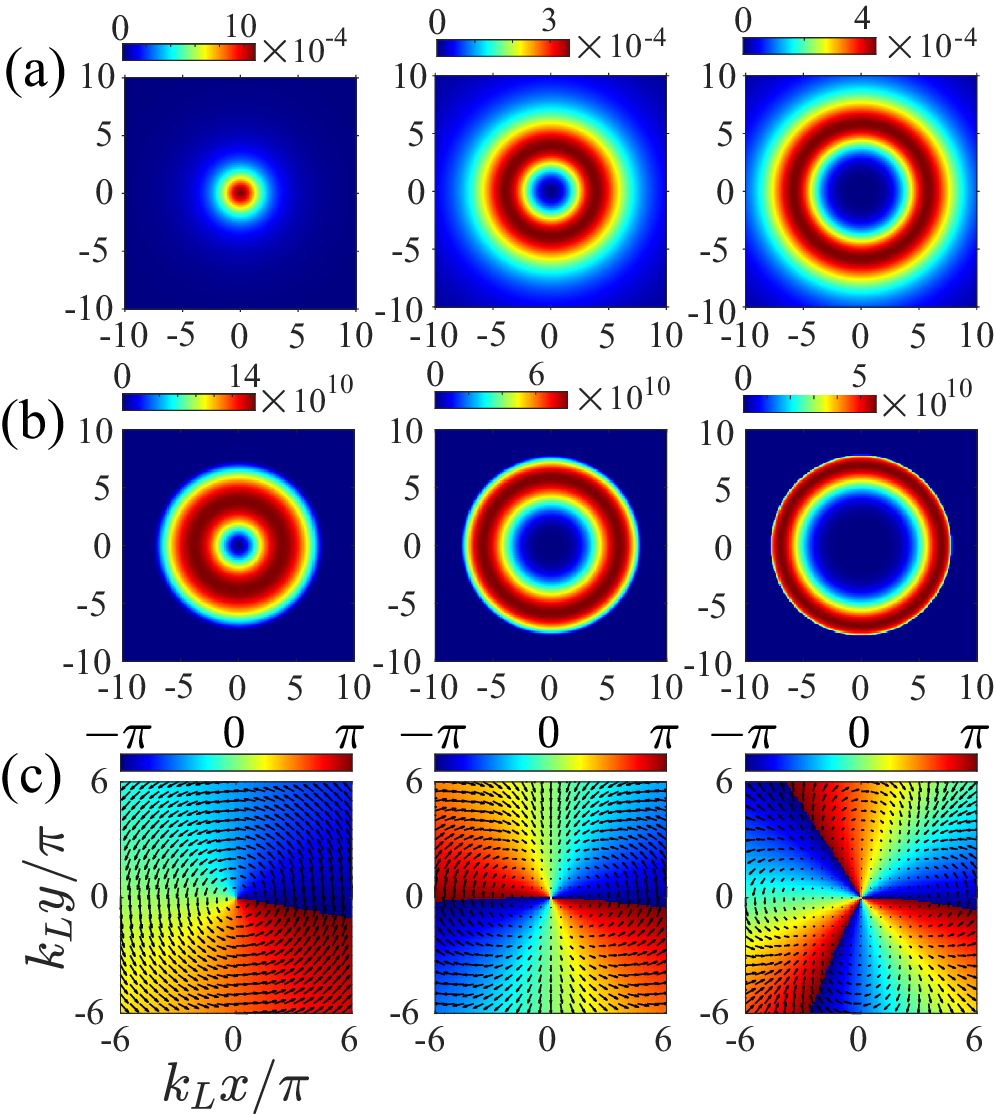} 
	\caption{(a) SOAMC-mediated spatial distribution of a synthetic magnetic field ${\cal B}(\vartheta=0)$. (b) The typical density distribution of the spin-$\downarrow$ atom $\rho_{\downarrow}=|\Psi_{\downarrow}|^2$ (units of cm$^{-2}$). (c) The corresponding phase $\phi_{\downarrow}={\rm arg}(\Psi_{\downarrow})$ and planar spin $\bm{S}_{\perp}$. The color with the blue-red gradient shading indicates the values ${\cal B}(\vartheta=0)$, $\rho_{\downarrow}$, and $\phi_{\downarrow}$, respectively. From left to right, the OAM quantum numbers are $l =$1 to 3. The other parameters are ($\delta,\Omega_\mathrm{R}$)=(-5,12)$E_L/\hbar$ and  $\vartheta =0$.}
	\label{Fig2}
\end{figure} 
Remarkably, the synthetic magnetic field described in equation~\eqref{eq6} is solely dependent on the radial position $r$, and is independent of the polar angle $\varphi$. By performing a gauge transformation incorportating an arbitrary rotation angle $\varphi'$,  the spatial distribution of  ${\cal B}(\vartheta=0)$ remains invariant, thereby preserving a continuous rotational symmetry in the presence of SOAMC.

Figure~\ref{Fig2}(a) shows the synthetic magnetic field ${\cal B} (\vartheta=0)$ generated by SOAMC for a varying angular quantum number $l$ with $\delta = -5 E_{\mathrm L}/\hbar$ and $\Omega_\mathrm{R} = 12 E_{\mathrm L}/\hbar$. For an angular quantum number $l=1$ of the LG laser, the spatial distribution of ${\cal B}(\vartheta=0)=\frac{\hbar C}{2(1+Cr^2)^{{3}/{2}}}\bm{\hat{e}_z}$  forms a circular pattern, with the maximum amplitude locating at the origin with the value ${\hbar C}/{2} $. With higher angular quantum numbers ($l>1$), the synthetic magnetic field is distributed in a radial doughnut shape~\cite{PhysRevLett.121.250401}, where the value of ${\cal B}(\vartheta=0)$ is zero at the center of the BEC and  the maximum value for the radial position is at $r_{\rm max}= [\frac{2l-2}{C (l+2)}]^{{1}/{2l}}\varpropto {(\delta/\Omega_\mathrm{R})}^{1/l}$ corresponding to ${\cal B}_{\rm max} \varpropto {(\Omega_\mathrm{R}/\delta)}^{2/l}$. As can be seen, both the inner and outer diameters of the ring expand as the angular quantum number $l$ increases. It is important to note that the magnitude of the synthetic magnetic field produced by SOAMC is approximately of the order of a few milliteslas, even for a significantly large laser waist $w_0 \gg 1/k_L$.

To gain a deeper understanding of the underlying physics, we calculate the ground-state structures of the condensate influenced by SOAMC. In figure~\ref{Fig2}(b), we display the typical density distribution $\rho_{\downarrow}=|\Psi_{\downarrow}|^2$ for the spin-$\downarrow$ atom. The wavefunction $\Psi_{\downarrow}$ is not shown since it does not host any interesting ground-state structures in our simulations. Of particular interest, $\rho_{\downarrow}$ demonstrates transitions from singly to triply quantized vortices induced by SOAMC, where the photon's OAM is transferred to the atomic wavefunction with respect to spin-flip of the internal states. The density profile of the multiply quantized vortex phase mirrors the spatial configuration of the synthetic magnetic field, maintaining the rotational symmetry. Unlike the typically unstable multiply quantized vortices, which tend to fragment into multiple singly quantized vortices~\cite{PhysRevA.91.043605,PhysRevA.77.053612}, our findings underscore that the SOAMC-mediated condensate provides a unique platform to achieve stable spin vortex states with a high angular quantum number against splitting. It should be noted that the observed vortex states are stable and robust against the variation of the $s$-wave collisional interaction.

\begin{figure}[h]
	\centering
	\includegraphics[width=0.59\textwidth]{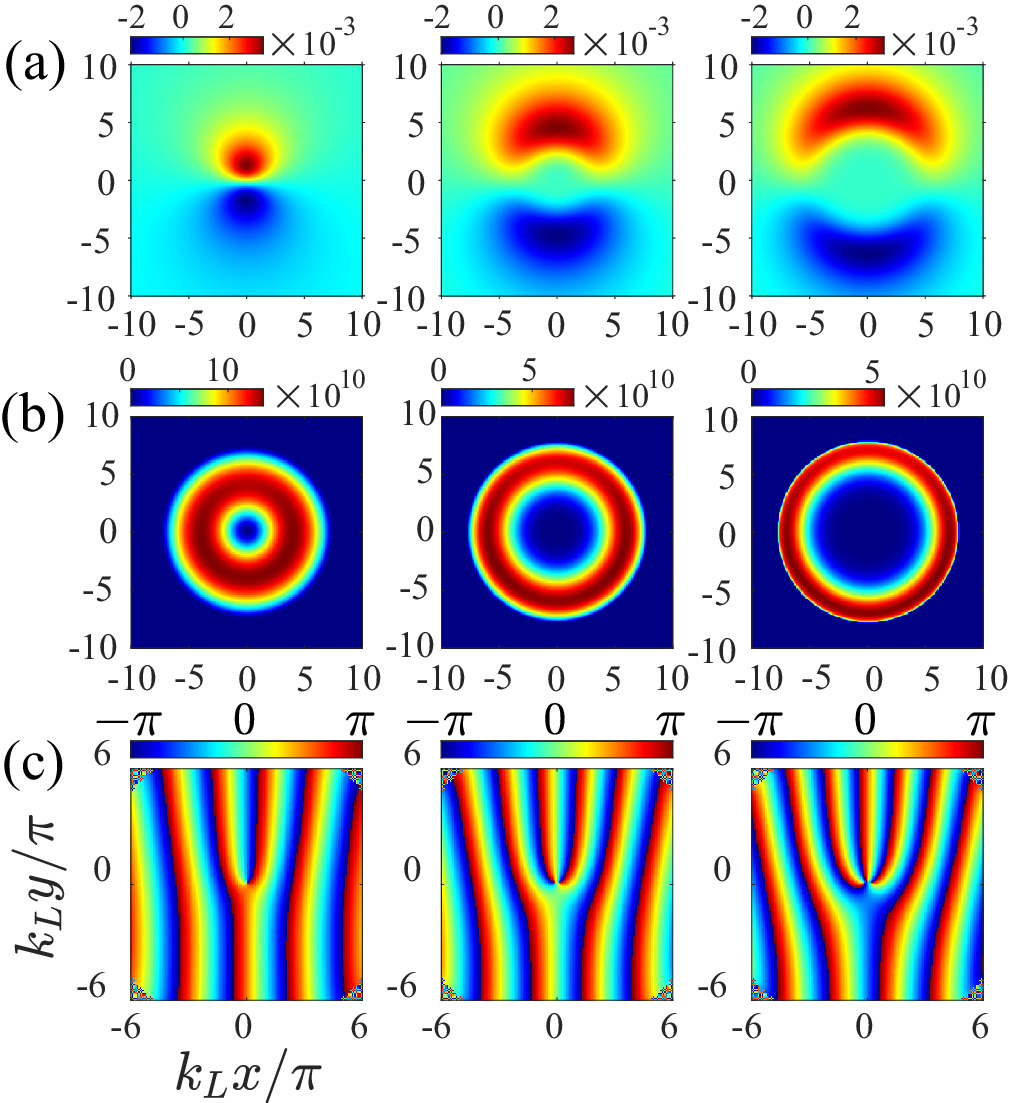} 
	\caption{(a) The spatial distribution of the synthetic magnetic field ${\cal B}(\vartheta)$ as a result of interplay of SLMC and SOAMC. (b) The density distribution $\rho_{\downarrow}$ (units of cm$^{-2}$). (c) The corresponding phase $\phi_{\downarrow}={\rm arg}(\Psi_{\downarrow})$.  The color with the blue-red gradient shading indicates the values ${\cal B}(\vartheta)$, $\rho_{\downarrow}$, and $\phi_{\downarrow}$, respectively. From left to right, the OAM quantum numbers are $l =$1 to 3. The other parameters are ($\delta,\Omega_\mathrm{R}$)=(-5,12)$E_L/\hbar$ and $\vartheta =\pi/2$.  }	
	\label{Fig3}
\end{figure} 

As expected, the winding number of the vortex phase~\cite{PhysRevLett.83.2498,PhysRevLett.93.250406,RevModPhys.59.533,RevModPhys.71.S318,RevModPhys.82.2785} matches the angular quantum number $l$ carried by the LG laser, as displayed in figure~\ref{Fig2}(c). Clearly, the vortex core size increases with the higher quantized winding number $l$. By introducing the pseudo-spin ${\bm S}({\bm r})=\sum_{\sigma\sigma'}\Psi_{\sigma}^* \bm{\hat{\sigma}}_{\sigma\sigma'}\Psi_{\sigma'}$ with $\bm{\hat{\sigma}}$ being the Pauli matrices, the structure of the synthetic magnetic field ${\cal B}$ also agrees with that of the  configuration of the planar spin $\bm{S}_{\perp}=(S_x,S_y)$. The vector plot in figure~\ref{Fig2}(c) displays the numerical results of
planar spin $\bm{S}_{\perp}$ by solving the Gross-Pitaevskii equations, and reveals a quadrupole spin texture characterized by a winding number of $-l$. This inverse correlation arises because $\bm{S}_{\perp}$ is antiparallel to the local field generated through Raman coupling mediated by the LG laser, which imparts a quantized OAM of $\hbar l$ to the atoms.

\section{Exotic vortex phase emerging as a result of SOAMC and SLMC}\label{four}
Now, we turn on the synthetic gauge field and exotic vortex with the combination of both SOAMC and SLMC. For $\sin\vartheta \neq 0$, a conventional SLMC with equal Rashba and Dresselhaus contributions~\cite{Lin2011} can be created from the single-particle Hamiltonian as shown in equation~\eqref{eq2}. After the gauge transformations $|\uparrow\rangle \rightarrow \mathrm{e}^{\mathrm{i}kx/2} |\uparrow\rangle$ and  $|\downarrow\rangle \rightarrow \mathrm{e}^{-\mathrm{i}kx/2} |\downarrow\rangle$~\cite{Goldman_2014,PhysRevLett.102.130401}, this mechanism introduces an additional vector potential ${\bm A}=-\hbar {k} \bm{\hat{\sigma_z}}/2$  without adiabatic approximation. Here, the wave vector of the light $k=k_L \sin{\vartheta}$ characterizes the strength of the one-dimensional SOC along the $x$ direction.
Unlike the rotational symmetry preserved under SOAMC, the interplay of SOAMC and SLMC gives rise to rotational symmetry breaking. Under the adiabatic approximation~\cite{RevModPhys.83.1523}, the synthetic magnetic field arising from the simultaneous presence of both SOAMC and SLMC can be derived as
\begin{equation} \label{eq7}
	\begin{split}
		{\cal {B}} (\vartheta)&= \frac{ \hbar  C }{2} (1+ C r^{2l})^{-\frac{3}{2}}\\ \times (l&+k_L \sin{\vartheta} \: y) l r^{2l-2}  \bm{\hat{e}_z}.
	\end{split} 
\end{equation}
Clearly, the strength and spatial modulation for ${\cal {B}}(\vartheta)$ can be significantly tuned by the SLMC of the running wave laser. Compared to ${\cal {B}}(\vartheta=0)$ in equation~(\ref{eq6}), the difference of synthetic magnetic field is satisfying 
\begin{equation} \label{eq8}
	\begin{split}
		\Delta{\cal {B}}={\cal {B}} (\vartheta)&- {\cal {B}}(\vartheta=0)\\
		=\frac{ \hbar k_Ly\sin{\vartheta} C }{2} (1&+ C r^{2l})^{-\frac{3}{2}} l r^{2l-2}  \bm{\hat{e}_z},
	\end{split} 
\end{equation}
The resulting anisotropic $\Delta{\cal {B}}$, driven by the SLMC, effectively disrupts the spatial rotational symmetry.

Figure~\ref{Fig3} shows the synthetic magnetic field and the ground-state density distribution as a result of the interplay of SOAMC and SLMC with the fixing of $\vartheta =\pi/2$. This configuration reveals a spatially uneven distribution of the synthetic magnetic field, ${\cal B}(\vartheta)$, characterized by broken rotational symmetry. This is a marked departure from the doughnut-shape distribution created by SOAMC alone. Of particular interest, we find that the tensity of ${\cal B}(\vartheta)$ is approximately an order of magnitude higher  than that observed in figure~\ref{Fig2}, suggesting a significantly enhanced mechanism for generating synthetic magnetic fields with the combination of SLMC and SOAMC. With regard to the condensate density distribution, the observed vortex phase with the winding number $l$ also hosts rotational asymmetric density distribution (figure~\ref{Fig3}(b)). This anisotropy mirrors the uneven spatial profile of ${\cal B}(\vartheta)$, further evidencing the breaking of rotational symmetry. Intriguingly, we identify a complex phase structure within the multiply quantized vortex phase, as shown in  figure~\ref{Fig3}(c). The emergence of phase stripes with multiply-quantized singularity plots of the wavefunction is due to the interference between SOAMC and SLMC. Upon applying the SLMC ($\vartheta>0$), the phase of the ground state will undergo the phase transition immediately, corresponding to the rotational symmetry breaking. This rotational asymmetry in the phase profile is distinct from the phase stripes generated by one-dimensional SLMC in cold atom experiments~\cite{Lin2011,li2017stripe}, underscoring the unique interplay between SOAMC and SLMC in our observations.

\begin{figure}[h]
	\centering
	\includegraphics[width=0.65\textwidth]{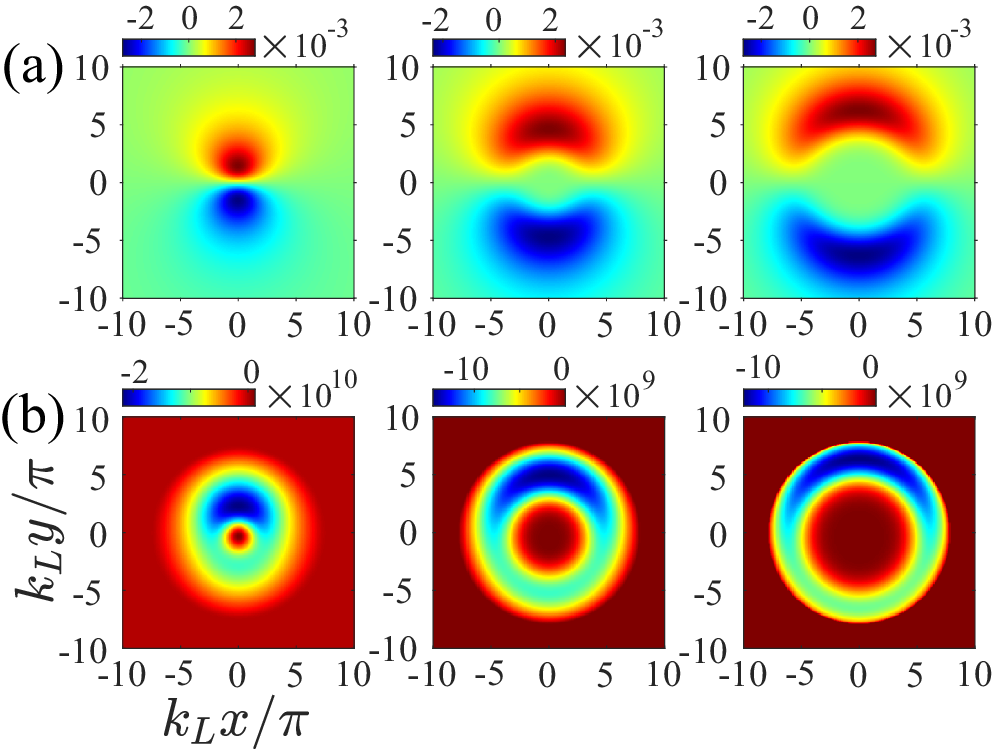} 
	\caption{(a) The different synthetic magnetic field $\Delta{\cal B}$ and (b) density distribution $\delta \rho_{\downarrow}$(units of cm$^{-2}$) as a result of interplay of SLMC and SOAMC between the relative angle $\vartheta =\pi/2$ and  $\vartheta =0$. From left to right, the OAM quantum numbers are $l =$1 to 3. The other parameters are $(\delta,\Omega_\mathrm{R})=(-5,12)E_L/\hbar$.}
	\label{Fig4}
\end{figure}

To delve deeper into the phenomenon of rotational symmetry breaking, we present the difference synthetic magnetic field $\Delta{\cal B}$ and the variation in density distribution $\delta \rho_{\downarrow}$ for the vortex phase when comparing the relative angle $\vartheta =\pi/2$ and  $\vartheta =0$  in figure~\ref{Fig4}. As shown in the plots, the effect of SLMC on the synthetic magnetic field $\Delta{\cal B}$ and density distribution $\delta \rho_{\downarrow}$ can be clearly extracted. It is evident that the amplitude of $\Delta{\cal B}$ preserves the mirror symmetry along the $x$ axis, yet exhibits mirror asymmetry along the $y$ axis. This result can be well understood from the analytical expression equation~\eqref{eq8}. Indeed, this mirror symmetry results from the interplay between the rotational symmetry inherent to SOAMC and the
translational symmetry along the $x$ axis introduced by SLMC. Remarkably, the density difference $\Delta \rho_{\downarrow}=\rho_{\downarrow} (\vartheta=\pi/2) -\rho_{\downarrow} (\vartheta = 0)$ also exhibits the same mirror symmetry along the $x$ axis. The density pattern of $\Delta \rho_{\downarrow}$ shows a similar structure for the anisotropic synthetic magnetic field, where the positive value of $\Delta{\cal B}$ corresponds to a large density difference of $ |\Delta \rho_{\downarrow}|$. Our result underscore the critical role of the synthetic magnetic field in shaping the ground-state wavefunction of the condensate.

\section{Summary}\label{five}
Based on state-of-the-art SOC in ultracold quantum gases, we propose an experimental scheme to generate SOAMC and SLMC by utilizing a synergistic arrangement of a running wave and LG laser fields within spinor BECs. Subsequently, the synthetic magnetic field and ground-state structure of a Raman-dressed condensate have been investigated. In particular, the stable vortex state with a phase stripe hosting single multiply quantized singularity have been observed.
This discovery marks a departure from traditional observations of unstable multiply quantized vortices, which tend to fragment into single quantized vortices ~\cite{PhysRevA.102.063328,PhysRevResearch.2.033152}. The spin vortex states, characterized by their OAM and distinctive spin textures, can be experimentally measured using absorption images techniques. Our findings elucidate that the dynamic interplay between SOAMC and SLMC gives rise to the anisotropic synthetic magnetic field and the density pattern within the multiply quantized vortex phase, which is with respect to the rotational symmetry breaking but preserves the mirror symmetry. Remarkably, we point out that the rotational symmetry broken density distribution of the ground-state wavefunction is highly in agreement with the analytic results of the  anisotropic synthetic magnetic field in the dressed spin state.  Finally, the proposed scheme opens interesting opportunities for the exploration of exotic topological vortex phases and fundamental quantum mechanics in spin-orbital-angular-momentum coupled ultracold quantum gases. 

\section*{Acknowledgement}
{This work was supported by the NSFC (Grants No. 12274473 and No. 12135018).}




\end{document}